\documentclass[conference]{IEEEtran}

\ifCLASSINFOpdf
\else
\fi

\usepackage{url}
\usepackage[]{fancyhdr} %
\newcommand{\changefont}{\fontsize{9}{9}\selectfont}
\IEEEoverridecommandlockouts

\usepackage{amsmath, mathtools}
\usepackage{subfig}
\usepackage{longtable}
\usepackage{supertabular}
\usepackage{color}

\usepackage{scalefnt}
\usepackage{lipsum}

\begin{document}

\title{Day-ahead Schedule Considering the Participation of Electric Vehicles in Primary Frequency Response}

\author{\IEEEauthorblockN{Amanda Fernandes\IEEEauthorrefmark{1},
Miguel Carrión\IEEEauthorrefmark{2},
Ruth Domínguez\IEEEauthorrefmark{3},
Rodrigo Ramos\IEEEauthorrefmark{4}, 
Ahda Pavani\IEEEauthorrefmark{5} and
Werbeston Oliveira\IEEEauthorrefmark{1}}
\IEEEauthorblockA{\IEEEauthorrefmark{1} Federal University of Amapá, Amapá, Brazil}
\IEEEauthorblockA{\IEEEauthorrefmark{2} University of Castilla-La Mancha, Toledo, Spain}
\IEEEauthorblockA{\IEEEauthorrefmark{3} University of Brescia, Brescia, Italy}
\IEEEauthorblockA{\IEEEauthorrefmark{4} University of São Paulo, São Paulo, Brazil}
\IEEEauthorblockA{\IEEEauthorrefmark{5} Federal University of ABC, São Paulo, Brazil}}

\maketitle
\thispagestyle{fancy}
\pagestyle{fancy}

\begin{abstract}
The insertion of renewable sources in power systems may cause a decrease in the system's equivalent inertia, which result in the instability of the power system. On the other hand, energy storage systems have proven to be an effective tool to increase the flexibility in the operation of energy systems, which may favor integrating renewable energy sources. In this way, plug-in electric vehicle (PEVs) batteries can be used as storages when such vehicles are parked and connected to the grid. In this work, the contribution of PEVs to primary frequency response is analyzed in systems dominated by renewable sources. A day-ahead scheduling model is developed considering PEVs groups can actively participate in electricity markets, supporting day-ahead reserve capacity and providing primary frequency response. The proposed model is implemented in the distribution system of the Federal University of Amapá.
\end{abstract}

\begin{IEEEkeywords}
Energy Generating Schedule. Plug-in Electric Vehicles. Primary Frequency Regulation. Renewable Energy Source.
\end{IEEEkeywords}

\IEEEpeerreviewmaketitle

\section*{Notation}

    The notation used in this paper is included below.

\subsection{Indices/Set}

\begin{supertabular}{p{0.1\textwidth}p{0.78\textwidth}}
        $d/D$ & Index/set of consumers \\
        $D_{n}$ & Set of consumers located in bus $n$ \\
        $g/G$ & Index/set of generating units \\
        $G^{C/R}$ & Set of conventional/renewable generating \newline units \\
        $G_{n}$ & Set of generating units located in bus $n$ \\
        $k/K$ & Index/set of contingencies \\
        $l/L$ & Index/set of transmission line \\
        $L_{n}^{O/F}$ & Set of lines whose origin/destination bus is \newline $n$ \\
        $n/N$ & Index/set of buses \\
        $N^{O/F}_{l}$ & Set of origin/destination bus of line $l$ \\
        $t/T$ & Index/set of time periods \\
        $T_{v}$ & Set of periods in which PEVs in group $v$ \newline can be charged or discharged to the grid \\
        $v/V$ & Index/set of PEV groups \\
    \end{supertabular}
    
\subsection{Variables}    
     
    \begin{supertabular}{p{0.1\textwidth}p{0.78\textwidth}}    
        $c_{gt}^{SU/SD}$ & Startup/shutdown cost of generator $g$ in \newline period $t$ \\
        $c_{t}^{P}$ & Operation cost in period $t$ \\
        $c_{t}^{SP}$ & Spillage cost of renewable units in period $t$ \\ 
        $c_{t}^{SU/SD}$ & Startup/shutdown cost in period $t$ \\
        $c_{t}^{UD}$ & Unserved demand cost in in period $t$ \\
        $c_{t}^{UD,PR}$ & Unserved demand cost in period $t$ after \newline contingency $k$ \\
        $cc_{t}^{V}$ & Frequency reserve schedule by PEVs group \newline cost in period $t$ \\
        $cp_{t}^{V}$ & Deployed energy cost in period $t$ \\
        $c_{t}^{\Delta f}$ & Frequency deviation cost in period $t$ \\
        $c_{vnt}^{V,PR}$ & Scheduled capacity that can be used for PFR \newline by PEVs group $v$ in bus $n$ and period $t$ \\
        $e_{vnt}^{C/D}$ & Energy charged/discharged by PEVs group $v$ \newline in bus $n$ and period $t$ \\
        $e_{vntk}^{C/D,PR}$ & Energy charged/discharged by PEVs group $v$ \newline in PFR in bus $n$, period $t$ and after a \newline  contingency $k$ \\
        $e_{vntk}^{V}$ & Energy stored by PEVs group $v$ in bus $n$, \newline period $t$ and after a contingency $k$ \\
        $p_{dt}^{UD}$ & Unserved demand of consumer $d$ in period $t$ \\
        $p_{dtk}^{UD,PR}$ & Unserved demand of consumer $d$ in period $t$ \newline after contingency $k$ \\
        $p_{gt}$ & Power produced by unit $g$ in period $t$ \\
        $p_{lt}^{L}$ & Power flow in line $l$ in period $t$ \\
        $p_{vntk}^{V,PR}$ & Primary frequency response provided by PEVs \newline in group $v$ in bus $n$, period $t$ and after a \newline contingency $k$ \\
        $p_{vntk}^{V,PRC}$ & Primary frequency response provided by PEVs \newline in group $v$ in charging mode in bus $n$, period \newline $t$ and after a contingency $k$ \\
        $p_{vntk}^{V,PRD}$ & Primary frequency response provided by PEVs \newline in group $v$ in discharging mode in bus $n$, \newline period $t$ and after a contingency $k$ \\
        $p_{gtk}^{PR}$ & Primary reserve output of generator unit $g$ in \newline period t after contingency $k$ \\
        $s_{gt}$ & Spillage of renewable unit $g$  in period $t$ \\
        $u_{gt}$ & Binary variable that is equal to 1 if the unit \newline $g$  is online, and 0 otherwise \\
        $\Delta f_{tk}$ & Frequency deviation in period $t$ after \newline contingency $k$ \\
    \end{supertabular}
    
\subsection{Parameters}    
    \begin{supertabular}{p{0.1\textwidth}p{0.78\textwidth}}
        $A_{gt}$ & Availability factor of renewable unit $g$ in \newline period $t$, varying between 0 and 1 \\
        $C_{g}$ & Production cost of generator $g$  \\
        $C_{g}^{SU/SD,F}$ & Startup/shutdown cost factor of generator $g$  \\
        $Cc_{vnt}^{V,PR}$ & PFR capacity cost offer of PEV group $v$ in \newline bus $n$ and period $t$ \\
        $Cp_{vnt}^{V,PR}$ & Deployed reserve cost offer of PEV group $v$ \newline in bus $n$, period $t$ and after contingency $k$\\
        $C^{P}$ & Penalization cost of forced intermittent \newline power spillage \\
        $C^{UD}$ & Penalization cost of unserved demand \\
        $C^{\Delta F}$ & Penalization cost of frequency deviation \\
        $D^{PR}$ & Maximum duration of primary frequency \newline response in hours \\
        $DR_{g}$ & Frequency droop of generator unit $g$  \\
        $DR_{v}$ & Frequency droop of PEV group $v$ \\
        $DT_{g}$ & Minimum down time of unit $g$  \\
        $E_{max,v}^{V}$ & Capacity of batteries of PEVs in group $v$ \\
        $E_{min,v}^{V}$ & Minimum value of energy that must \newline remain in the batteries of PEVs in group $v$ \newline in each charging/discharging period \\
        $E_{vn}^{VF}$ & Minimum status of PEVs batteries in \newline group $v$ and period $t$ at the end of charging \newline period \\
        $E_{vn}^{V0}$ & Initial status of the batteries of PEVs in \newline group $v$ and period $t$ at the beginning of the \newline charging period \\
        $N_{vn}^{V}$ & Number of PEVs in group $v$ \\
        $P_{dt}^{D}$ & Power demand of supplied to consumer $d$ \newline in period $t$ \\
        $P_{max,g}$ & Maximum power output of unit $g$  \\
        $P_{max,lt}^{L}$ & Capacity of line $l$ in period $t$ \\
        $P_{max}^{V}$ & Maximum power charging/discharging rate \newline of PEVs \\
        $P_{min,g}$ & Minimum power output of unit $g$  \\
        $PR_{g}^{U/D}$ & Ramp-up/down limit of unit $g$  \\
        $S_{gk}$ & Availability parameter, equal to 1 if the unit \newline $g$ is out after contingency $k$ and 0 otherwise \\
        $TC_{g}$ & Nº of periods that unit $g$ must be initially \newline offline \\
        $TG_{g}$ & Nº of periods that unit $g$  must be initially \newline online \\
        $T_{v}^{0/F}$ & Initial/final period which PEVs group $v$ can \newline be charged or discharged \\
        $UT_{g}$ & Minimum up time of unit $g$  \\
        $X_{l}$ & Reactance of line $l$ \\
        $\eta_{v}$ & Efficiency of PEVs in group $v$\\
    \end{supertabular}

\section{Introduction}

    The presence of renewable energy sources (RES) in electrical power systems has increased due to environmental concerns. The emission of Greenhouse Gases is one of the biggest concerns of our society due to global warming \cite{Paris, ODS}. The electricity production sector is one of the main emitters of carbon dioxide ($CO_2$). For this reason, there is a great need for the increase of carbon-free sources in the electrical system to reduce the use of conventional fossil energy \cite{Irena}.
    
    In this context, the Federal University of Amapá (UNIFAP) started in August 2020 the installation of a photovoltaic solar plant, as part of the project "UNIFAP SOLAR: Implantação de Geração Distribuída Fotovoltaica no Campus Marco Zero do Equador". This project aims to implement environmental sustainability programs at the university, as well as promote actions for the institutional community through the reduction and reuse of resources and energy \cite{unifap-solar}. The installation that has started foresees the installation of about 1.3 MWp in solar photovoltaic power plants at UNIFAP. 
    
    Despite the efforts to increase generation through renewable sources, it can be seen that the presence of renewable sources in isolated systems has occurred on a smaller scale than in interconnected systems \cite{Carrion-2019}. One of the main reasons is that the intermittency of energy resources such as solar and wind generates energy quality problems. The insertion of renewable sources in the system causes a decrease in the system's equivalent inertia, which may result in threads of instability.

    In contrast, the use of energy storage systems has proven to be an effective tool to increase the flexibility of the energy system operation \cite{Dunn-2011}, facilitating the installation of renewable sources in isolated power systems. The Vehicle-to-Grid ($V2G$) capability of Plug-in Electric Vehicles (PEVs), allows vehicles to inject the energy stored in the batteries into the grid. As a consequence of this, PEVs can provide ancillary services to the electric power system when connected to the grid. To balance power consumption and production, PEVs can be used as both a load and a generating sources to maintain the system frequency at acceptable values by charging their batteries when there is too much generation in the grid and acting as generators by discharging the batteries when there is too much load in the system \cite{Kempton-2008}.
    
    Despite the numerous advantages of renewable technologies, the production coming from these sources can be considered as non-dispatchable, variable, and uncertain. Traditional synchronous generators can provide inertia and primary frequency response (PFR) to power systems. However, renewable energy is connected to the grid by the power converters, which are unable to provide inertia to the system in a similar manner than synchronous generators \cite{Fanglei2020}. In a context in which renewable units are supplying a great part of the demand, the determination of the day-ahead scheduling is more important and complex than in thermal-dominated power systems. The day-ahead schedule is a large-size mixed-integer linear programming (MILP), the determination of the day-ahead scheduling of a power system is a complex mathematical problem that is based on the formulation of the economic dispatch or unit commitment problems.
    
    The insertion of renewable energy sources and electric vehicles in power systems has been extensively studied in recent years. In \cite{Restrepo-2005} the UC that simultaneously accounts for both primary and tertiary reserve constraints is formulated. In Reference \cite{Carrion-2006} a mixed-integer linear formulation for the thermal UC problem is presented in order to reduce the computational burden of existing MILP approaches. A two-stage stochastic UC model with high renewable penetration is present in \cite{asensio2015stochastic}, the model proposed is applied in the power system of the Canary Islands and Crete. 
    
    Reference \cite{mercier} studies the support of the battery energy storage system in the dynamic stability of an isolated power system with low grid inertia. The participation of electric vehicles providing frequency control is examined in \cite{almeida2011electric}. In \cite{Yoo2019} the frequency-support parameters of energy storage systems are calculated for achieving stable frequency response from a power system with high penetration of renewable generators. An economic feasibility study of V2G frequency regulation is performed in \cite{han2013economic}, in consideration of battery wear. Reference \cite{zhang2016day} propose an algorithm that optimizes the charging/discharging of energy storage devices in order to minimize the total system day-ahead operating cost.
    
    Reference \cite{aziz2018electric} presents a study of electric vehicle utilization to support a small-scale energy management system, showing that the utilization is feasible and deployable. In Reference \cite{bellekom2012electric}, the Dutch power system operation is analyzed when the penetration of EV and RES is expected to increase significantly due to the goals to reduce CO2 emissions. In \cite{carrion2015operation} the participation of PEVs in a renewable-dominated power system based on the isolated power system of Lanzarote-Fuerteventura, Spain, is analyzed.
    
    In this context, considering the need of introducing carbon-free technologies in the energy system, this work will assess the participation of PEVs in the day-ahead energy generation market and the provision of ancillary services in a predominantly renewable isolated system. In this work, the mathematical modeling is based on \cite{Carrion-2019}, with some simplifications for a less complex system. The microgrid to be used in this work corresponds to the electrical system of the Marco Zero Campus of the Federal University of Amapá. For the development of the analysis, it will consider that it is a system connected to the grid with isolated operation capacity, with about $40\%$ of the energy mix being renewable sources.

\section{Model}

    A linear programming formulation is used in this work, where the objective function to be minimized is the operating cost of an electrical power system. Among the restrictions to which the objective function is subject, stand out those constraints related to PEVs, renewable energy sources, load flow, frequency deviation, frequency regulation, and operation of the generating units.

\subsection{Problem Formulation}
\label{section-equations}

    The formulation described in this work was based in \cite{Carrion-2019}. The mathematical formulation of the proposed unit commitment problem is the following:

    {
    \fontsize{7pt}{\baselineskip}\selectfont
    \begin{equation}
        \begin{split}
        \MoveEqLeft
           \emph{Min} \hskip 0.2cm \sum_{t \in T} \sum_{g \in G} (C_{g}\cdot p_{gt} + c_{gt}^{SU} + c_{gt}^{SD}) + \sum_{t \in T} \sum_{d \in D} C^{UD} \cdot p_{dt}^{UD} \\
           &+ \sum_{t \in T} \sum_{g \in G^R} C^{P} \cdot s_{gt}
            +  \sum_{d \in D} \sum_{t \in T} \sum_{k \in K} (C^{UD} \cdot p_{dtk}^{UD,PR} -  C^{\Delta F} \cdot \Delta f_{tk}) \\ 
            &+  \sum_{t \in T} \sum_{n \in N} \sum_{v \in V} (Cc_{vnt}^{V,PR} \cdot c_{vnt}^{V,PR}
            + \sum_{k \in K} Cp_{vnt}^{V,PR} \cdot p_{vntk}^{V,PR})
        \end{split}
        \label{objfunction}
    \end{equation}
    }
    
    Subject to:
    
    {
    \fontsize{7pt}{\baselineskip}\selectfont
        \begin{equation}
        \begin{split}
        \MoveEqLeft
            \sum_{g \in G_{n}} p_{gt} + \sum_{l \in L_{n}^{F}} p_{lt}^{L} - \sum_{l \in L_{n}^{O}} p_{lt}^{L} + \sum_{v \in V} (e_{vnt}^{D}-e_{vnt}^{C}) \\
            &= \sum_{d \in D_{n}} (p_{dt}^{D} - p_{dt}^{UD}),
            \hskip 0.3cm \forall t \in T, \forall n \in N
            \end{split}
            \label{energybalance}
        \end{equation}

        \begin{equation}
            - P_{max,lt}^{L} \leq p_{lt}^{L} \leq P_{max,lt}^{L},
            \hskip 0.3cm \forall l \in L, \forall t \in T
            \label{flowlimits}
        \end{equation}
        
        \begin{equation}
            p_{lt}^{L} = \frac{1}{X_{l}} \cdot (\theta_{N^{O}_{lt}}- \theta_{N^{F}_{lt}}),
            \hskip 0.3cm \forall l \in L, \forall t \in T
            \label{flowline}
        \end{equation}

        \begin{equation}
            P_{min,g} \cdot u_{gt} \leq p_{gt} \leq P_{max,g} \cdot u_{gt},
            \hskip 0.3cm \forall g \in G^C, \forall t \in T
            \label{limitsC}
        \end{equation}
    
        \begin{equation}
            p_{gt} + s_{gt} = P_{max,g} \cdot A_{gt},
            \hskip 0.3cm \forall g \in G^R, \forall t \in T
            \label{limitsR}
        \end{equation}

        \begin{equation}
            p_{gt} - p_{g t-1} \leq PR_{g}^{U},
            \hskip 0.3cm \forall g \in G^C, \forall t \in T
            \label{prampup}
        \end{equation}

        \begin{equation}
            p_{gt-1} - p_{gt} \leq PR_{g}^{D}
            \hskip 0.3cm \forall g \in G^C, \forall t \in T
            \label{prampdown}
        \end{equation}

        \begin{equation}
            c_{gt}^{SU} \geq C_{g}^{SU,F} (u_{gt}-u_{gt-1}),
            \hskip 0.3cm \forall g \in G^C, \forall t \in T
            \label{startup}
        \end{equation}
    
        \begin{equation}
            c_{gt}^{SD} \geq C_{g}^{SD,F} (u_{gt-1}-u_{gt}),
            \hskip 0.3cm \forall g \in G^C, \forall t \in T
            \label{shutdown}
        \end{equation}
    
        \begin{equation}
            c_{gt}^{SU} \geq 0,  c_{gt}^{SD} \geq 0 \hskip 0.3cm \forall g \in G^{C}, \forall t \in T
            \label{startup-shutdown}
        \end{equation}

        \begin{equation}
            \sum_{t=1}^{TG_{g}} (1-u_{gt}) = 0,
            \hskip 0.3cm \forall g \in G^C
            \label{uptime01}
        \end{equation}
    
        \begin{equation}
            \begin{split}
                &\sum_{\tau =t}^{t-UT_{g}-1} u_{g\tau} \geq UT_{g}(u_{gt} - u_{gt-1}),\\
                 \forall g& \in G^C,  \forall t = TG_g + 1 \cdots T-UT_g+1
            \end{split}
            \label{uptime02}
        \end{equation}
    
        \begin{equation}
            \begin{split}
            &\sum_{\tau=t}^{T} [u_{g\tau} - (u_{gt}-u_{gt-1})] \geq 0, \\
                 \forall &g \in G^C, \forall t = T - UT_g +2 \cdots T
            \end{split}
            \label{uptime03}
        \end{equation}

        \begin{equation}
            \sum_{t=1}^{TL_{g}} u_{gt} = 0,
            \hskip 0.3cm \forall g \in G^C
            \label{downtime01}
        \end{equation}
        
        \begin{equation}
            \begin{split}
                &\sum_{\tau=t}^{t+DT_g-1} (1-u_{g\tau}) \geq DT_g(u_{gt-1} - u_{gt}), \\
                & \forall g \in G^C, \forall t = TL_g+1 \cdots T-DT_g+1
            \end{split}
            \label{downtime02}
        \end{equation}
    
        \begin{equation}
            \sum_{\tau=t}^{T} [1-u_{g\tau}-(u_{gt-1}-u_{gt})] \geq 0,
            \hskip 0.3cm \forall g \in G^C, \forall t = T-DT_g+2 \cdots T
            \label{downtime03}
        \end{equation}

        \begin{equation}
            0 \leq p_{gtk}^{PR} \leq - \frac{1}{DR_{g}} \cdot \Delta f_{tk},
            \hskip 0.3cm \forall g \in G^C \hskip 0.2cm | \hskip 0.2cm g \notin S_{gk}, \forall t \in T, \forall k \in K
            \label{pr01}
        \end{equation}
    
        \begin{equation}
            p_{gtk}^{PR} + p_{gt} \leq P_{max,g} \cdot u_{gt},
            \hskip 0.3cm \forall g \in G^C \hskip 0.2cm | \hskip 0.2cm g \notin S_{gk}, \forall t \in T, \forall k \in K
            \label{pr02}
        \end{equation}
    
        \begin{equation}
            p_{gtk}^{PR} = 0,
            \hskip 0.3cm \forall g \in G^{R} \hskip 0.2cm | \hskip 0.2cm g \notin S_{gk}, \forall t \in T, \forall k \in K
            \label{pr03}
        \end{equation}
        
        \begin{equation}
            p_{gtk}^{PR} = - p_{gt},
            \hskip 0.3cm \forall g \in G \hskip 0.2cm | \hskip 0.2cm g \in S_{gk}, \forall t \in T, \forall k \in K
            \label{pr04}
        \end{equation}
    
        \begin{equation}
            \sum_{d \in D} p_{dtk}^{UD,PR} + \sum_{g \in G} p_{gtk}^{PR} + \sum_{v \in V} \sum_{n \in N} p_{vntk}^{V,PR} = 0,
            \hskip 0.3cm \forall t \in T, \forall k \in K
            \label{pr05}
        \end{equation}

        \begin{equation}
            e_{vntk}^{V} = N_{vn}^{V} \cdot E_{vn}^{V0},
            \hskip 0.3cm \forall v \in V, \forall n \in N ,  t=T_{v}^{0}-1, \forall k \in K
            \label{statusbat01}
        \end{equation}
    
        \begin{equation}
            e_{vntk}^{V} \geq N_{vn}^{V} \cdot E_{vn}^{VF},
            \hskip 0.3cm \forall v \in V, \forall n \in N , t=T_{v}^{F}, \forall k \in K
            \label{statusbat02}
        \end{equation}
    
        \begin{equation}
            \begin{split}
            e_{vntk}^{V} = e_{vnkt-1}^{V} &+ \eta_{v} (e_{vnt}^{C} - e_{vntk}^{C,PR}) - \frac{1}{\eta_{v}} (e_{vnt}^{D} + e_{vntk}^{D,PR}),\\
                 \forall v &\in V, \forall n \in N , \forall t \in T_{v}, \forall k \in K
            \end{split}
            \label{energystoredpevs}
        \end{equation}
        
        \begin{equation}
            \begin{split}
            N_{vn}^{V} &\cdot E_{min,v}^{V} \leq e_{vntk}^{V} \leq N_{vn}^{V} \cdot E_{max,v}^{V}, \\
            & \forall v \in V, \forall n \in N , \forall t \in T_{v}, \forall k \in K
            \end{split}
            \label{limitsenergybat}    
        \end{equation}
        
        \begin{equation}
            e_{vnt}^{C}, e_{vnt}^{D} \leq N_{vn}^{V} \cdot P_{max,v}^{V},
            \hskip 0.3cm \forall v \in V, \forall n \in N , \forall t \in T, \forall k \in K
            \label{limitsenergybat2}  
        \end{equation}
        
        \begin{equation}
            e_{vntk}^{C,PR} = D^{PR} \cdot p_{vntk}^{V,PRC},
            \hskip 0.3cm \forall v \in V, \forall n \in N , \forall t \in T_{v}, \forall k \in K
            \label{pfrpev01}
        \end{equation}
        
        \begin{equation}
            D^{PR} \cdot p_{vntk}^{V,PRC} \leq e_{vnt}^{C} ,
            \hskip 0.3cm \forall v \in V, \forall n \in N , \forall t \in T_{v}, \forall k \in K
            \label{pfrpev02}
        \end{equation}
        
        \begin{equation}
            e_{vntk}^{D,PR} = D^{PR} \cdot p_{vntk}^{V,PRD},
            \hskip 0.3cm \forall v \in V, \forall n \in N , \forall t \in T_{v}, \forall k \in K
            \label{pfrpev03}
        \end{equation}
        
        \begin{equation}
            \begin{split}
            e_{vnt}^{D} &\leq N_{vn}^{V} \cdot P_{max}^{V} - D^{PR} \cdot p_{vntk}^{V,PRD},\\
             \forall v &\in V, \forall n \in N , \forall t \in T_{v}, \forall k \in K
            \end{split}
            \label{pfrpev04}
        \end{equation}
        
        \begin{equation}
            0 \leq p_{vntk}^{V,PRC} \leq N_{vn}^{V} \cdot P_{max}^{V},
            \hskip 0.3cm \forall v \in V, \forall n \in N , \forall t \in T_{v}, \forall k \in K
            \label{pfrpev05}
        \end{equation}
        
        \begin{equation}
            0 \leq p_{vntk}^{V,PRD} \leq N_{vn}^{V} \cdot P_{max}^{V},
            \hskip 0.3cm \forall v \in V, \forall n \in N , \forall t \in T_{v}, \forall k \in K
            \label{pfrpev06}
        \end{equation}
        
        \begin{equation}
            p_{vntk}^{V,PR}   = p_{vntk}^{V,PRC} + p_{vntk}^{V,PRD},
            \hskip 0.3cm \forall v \in V, \forall n \in N , \forall t \in T_{v}, \forall k \in K
            \label{pfrpev07}
        \end{equation}
        
        \begin{equation}
            0 \leq p_{vntk}^{V,PR}\leq - \frac{1}{DR_{v}} \cdot \Delta f_{tk},
            \hskip 0.3cm \forall v \in V, \forall n \in N , \forall t \in T_{v}, \forall k \in K
            \label{pfrpev08}
        \end{equation}
    
        \begin{equation}
            \begin{split}
                 p_{vntk}^{V,PR}&, e_{vnt}^{C}, e_{vnt}^{D}, e_{vntk}^{C,PR}, e_{vnt}^{D,PR} = 0, \\
                \forall v &\in V, \forall n \in N , \forall t \notin T_{v}, \forall k \in K
            \end{split}
            \label{pfrpev09}
        \end{equation}

        \begin{equation}
            0 \leq p_{vntk}^{V,PR}\leq c_{vnt}^{V,PR},
            \hskip 0.3cm \forall v \in V, \forall n \in N , \forall t \in T, \forall k \in K
            \label{pfrpev10}
        \end{equation}
        }
    \vskip-0.7cm
    The objective function (\ref{objfunction}) represents the expected costs considering production, startup and shutdown costs ($C_{g}\cdot p_{gt}$, $c_{gt}^{SU}$ and $c_{gt}^{SD}$, respectively)  and penalization for unserved energy ($C^{UD}\cdot p_{dt}^{UD}$), unserved primary reserve ($C^{UD} \cdot p_{dtk}^{UD,PR}$), frequency deviation ($C^{\Delta F} \cdot \Delta f_{tk}$), forced intermittent units spillage ($C^{P} \cdot s_{gt}$) and costs related to PEVs. Observe that the costs associated with frequency deviations and forced spillage are fictitious penalization costs intended to avoid frequency deviations and forced intermittent units spillage if possible.
    
    Constraint (\ref{energybalance}) presents the energy balance in the pre-contingency state. Constraints (\ref{flowlimits}) and (\ref{flowline}) represent the power flow in line $l$. The power limits of the generating units are presented by constraints (\ref{limitsC}) and (\ref{limitsR}). Constraints (\ref{prampup}) and (\ref{prampdown}) formulate the power ramps of generating units. Constraints (\ref{startup})-(\ref{startup-shutdown}) formulate startup and shutdown costs of generating units. Constraints (\ref{uptime01})-(\ref{uptime03}) represent the minimum up time of unit $g$ and the minimum down time is formulated by constraints (\ref{downtime01})-(\ref{downtime03}). The PFR is expressed by constraints (\ref{pr01})-(\ref{pr05}). Constraints (\ref{statusbat01})  and (\ref{statusbat02}) represent the status of batteries of PEVs. The energy stored by PEVs in each period $t$ is formulated by constraints (\ref{energystoredpevs}), (\ref{limitsenergybat}) and (\ref{limitsenergybat2}). Finally, the participation of PEVs in PFR is expressed by constraints (\ref{pfrpev01})-(\ref{pfrpev10}). Problem (\ref{objfunction})-(\ref{pfrpev10}) is a MILP problem that can be solved by commercial solvers.

\section{Case Study}    
\subsection{System Description}
\label{section-input}
    
    The microgrid used in this work corresponds to the Marco Zero Campus electrical system of the Federal University of Amapá. UNIFAP can be modeled as a consumer unit served at a voltage of 13.8 kV, with a contracted demand of 1000 kW off-peak and 1400 kW at peak hours, and an average monthly consumption of 341.780 kWh in 2019 \cite{diagnostico-pdp}.
    
    The microgrid comprises 64 buses, 63 lines, 32 consumer units, and 5 generators, where $G1$, $G2$ and $G3$ are conventional generators, and $G4$ and $G5$ are renewable energy sources. The technical characteristics of thermal and renewable generators are listed in Tables \ref{input-nonren-tab} and \ref{input-ren-tab}, respectively. Six PEV charging points were considered, which are located at strategic points of the university campus. In this work, three models of electric vehicles were used. Each model characterizes a group of vehicles. Groups 1 and 2 represent the PEVs  consisting on the academic community's own transport, whereas Group 3 corresponds to the bus belonging to the university's vehicle fleet. Table \ref{input-pevs-tab} provides the technical characteristics of PEVs.
    
    \begin{table*}[!h]
        \centering
        \caption{Technical characteristics of dispatchable units}
        \hfill\par
        \begin{tabular}{ccccccccccc}
            \hline & $C_{g}$ & $P_{max,g}$ & $P_{min,g}$ & $P0_{g}$ & $C_{gt}^{SU,F}$ & $C_{gt}^{SD,F}$ & $PR_{g}^{U}$ & $PR_{g}^{D}$ \\
            & (R\$/$MWh$) & ($MW$) & ($MW$) & ($MW$) & (R\$) & (R\$) & ($MW$) & ($MW$) \\ \hline
            Unit 1 & 505.0 &  0.60 &  0.12 &  0.30 &     909.00 &  9.09 &  0.15 &  0.15\\
            Unit 2 & 505.0 &  0.60 &  0.12 &  0.30 &     909.00 &  9.09 &  0.15 &  0.15\\
            Unit 3 & 505.0 &  0.60 &  0.12 &  0.30 &     909.00 &  9.09 &  0.15 &  0.15\\ \hline
        \end{tabular}
        \label{input-nonren-tab}
    \end{table*} 
    
    \begin{table}[!h]
        \centering
        \caption{Technical characteristics of intermittent units}
        \hfill\par
        \begin{tabular}{ccc}
            \hline
             & $C_{g}$ & $P_{max,g}$ \\
            & (R\$/$MWh$) & ($MW$)  \\ \hline
            Unit 4 & 0.000 & 0.554\\
            Unit 5 & 0.000 & 0.720\\ \hline
        \end{tabular}
        \label{input-ren-tab}
    \end{table} 

    \begin{table}[!h]
        \centering
        \caption{Technical characteristics of PEVs}
        \hfill\par
        \begin{tabular}{ccccc}
            \hline & $E_{max,v}^{V}$ & $E_{min,v}^{V}$ & $Cc_{v}^{V,PR}$ & $Cp_{v}^{V,PR}$  \\
             & ($MWh$) & ($MWh$) & (R\$/$MW$) & (R\$/$MWh$) \\ \hline
            Group 1 &  0.052& 0.0052&     50&    300\\
            Group 2 &  0.066& 0.0066&     50&    300\\
            Group 3 &  0.324& 0.0324&     50&    300\\ \hline
        \end{tabular}
        \label{input-pevs-tab}
    \end{table}
    
    The production costs of conventional generators included in Table \ref{input-nonren-tab}, $C_{g}$, were defined according to the ANEEL Tariff Ranking \cite{ranking} and it was considered that there is no production cost for intermittent sources. The cost of unserved demand is set at R\$10,000/$MWh$, aiming at a high penalty. The maximum allowed frequency deviation is considered 1 Hz.

\begin{figure}[!h]
        \centering
        \includegraphics[scale=0.4]{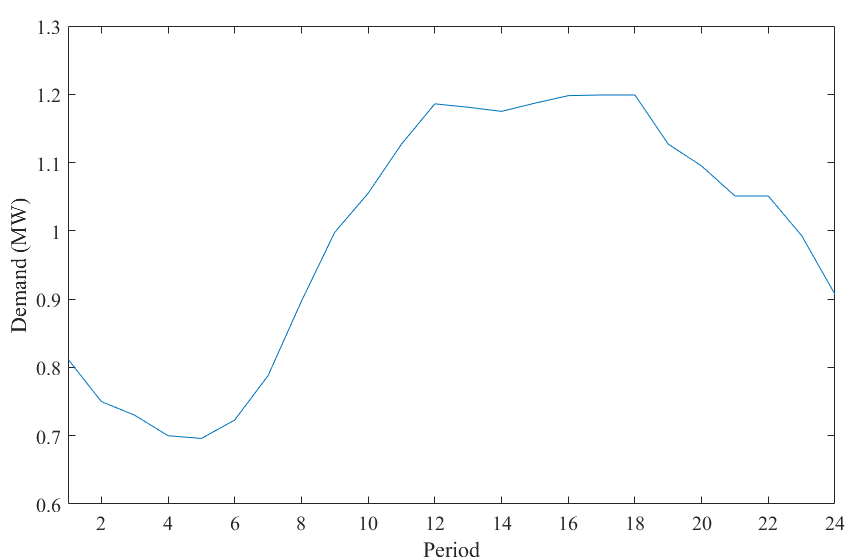}
        \caption{Energy demand in each period}
        \label{demand}
    \end{figure}

    The system demand in each period is shown in Figure \ref{demand}. It is important to highlight that the time period $t=1$ corresponds to the time 12 am, $t=2$ at 1 am, $t=3$ at 2:00, and so on until $t=24$ corresponding to 11 pm.
    
\begin{figure}[!h]
        \centering
        \includegraphics[scale=0.4]{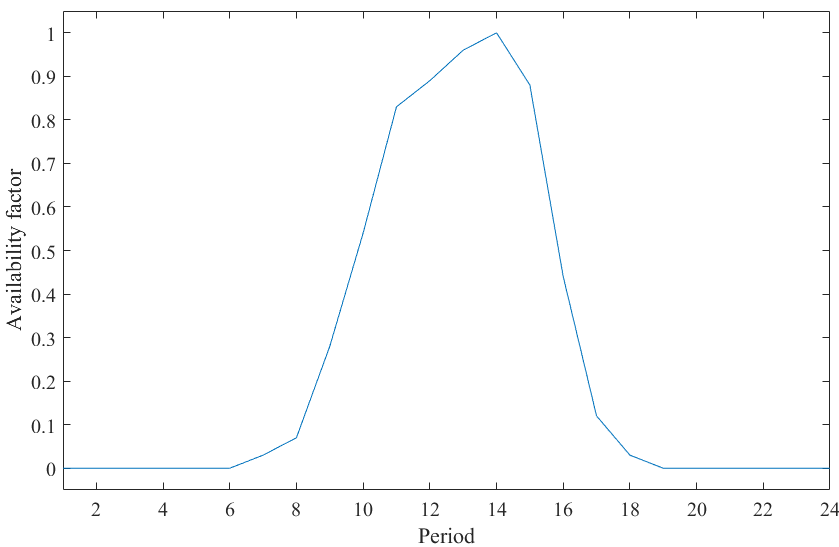}
        \caption{Availability factor of intermittent unit}
        \label{avail-factor}
    \end{figure}

    The availability factor of renewable units is presented in Figure \ref{avail-factor}, this value will limit the generation of these energy sources, that is, only in period 14 (when the availability factor is equal to 1) will the sources be able to generate the equivalent of their generation capacity.
    
\subsection{Results}
    
    The model proposed in Section \ref{section-equations} was tested on the system described in Section \ref{section-input}. All simulations are performed with GAMS using a laptop with $2.4$GHz processors and $4$GB of RAM.
    
    Three cases will be analyzed, these are:
    
    \begin{itemize}
        \item Case 1: Day-ahead scheduling without considering frequency reserve constraints;
        \item Case 2: Day-ahead scheduling with frequency reserve constraints. Only generation units participate in this service;
        \item Case 3: Day-ahead scheduling with frequency reserve constraints. Generation units and PEVs participate in this service;
    \end{itemize}
    
    \begin{figure*}[!h]
        \centering
        \includegraphics[scale=0.45]{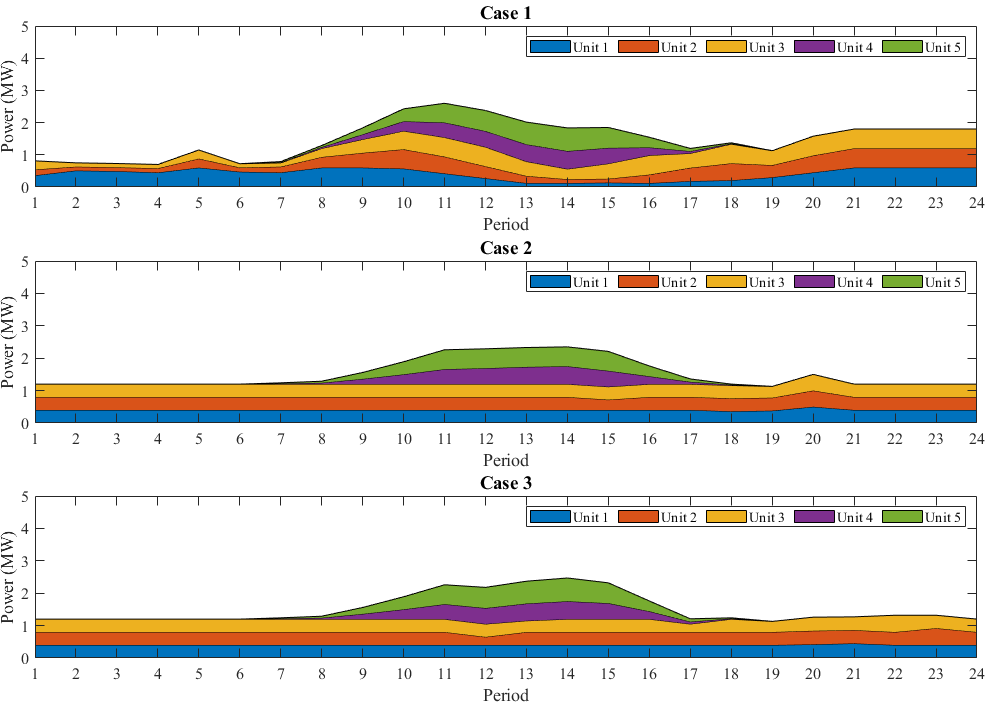}
        \caption{Power produced by units in each time period}
        \label{generation}
    \end{figure*} 
    
    Figure \ref{generation} shows the power produced by the generating units ($p_{gt}$) in the three considered cases. Please, observe that the first three units (Units $1$, $2$ and $3$) are conventional generating units, while the others (Units $4$ and $5$) are intermittent units. Note that when frequency regulation is not considered (Case 1), most of the energy produced in hours 1-7 is provided by Unit 1. Then, a failure of this unit may put at risk the operation of the system. However, if frequency regulation is considered, (Cases 2 and 3), the energy schedule considers the possibility of failures in the units and the demand in each hour is provided by several units. Observe that when units are operating at a low capacity level, and another unit fails, the low level unit cannot instantaneously increase to a high value of output power, due to ramp-up and ramp-down limits.
    
    \begin{figure*}[!h]
        \centering
        \includegraphics[scale=0.45]{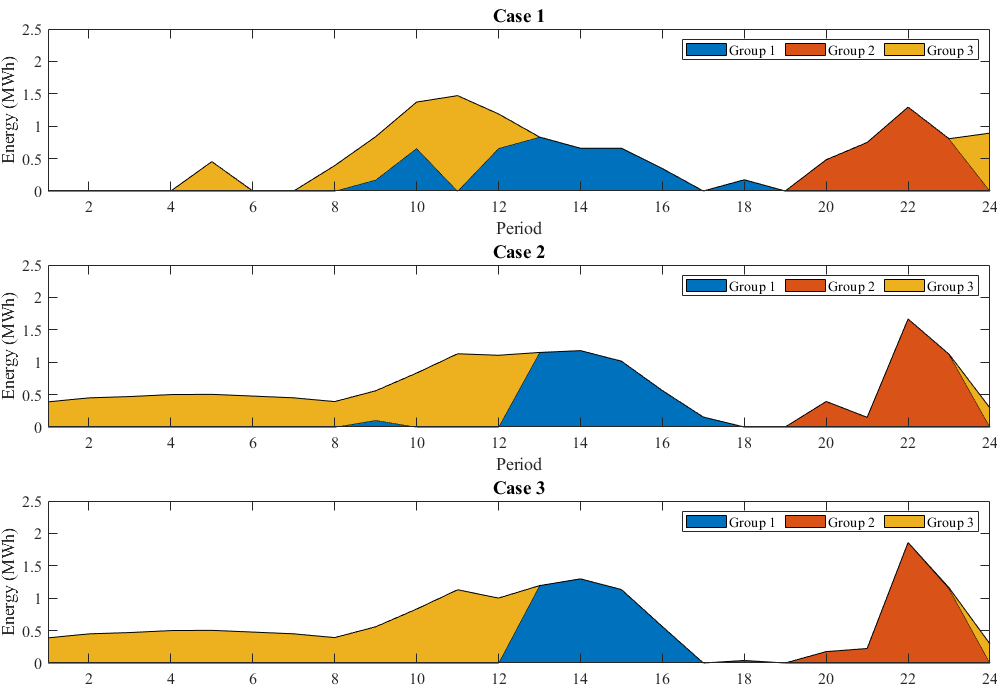}
        \caption{Energy charged by electric vehicles for each case}
        \label{energycharg-fig}
    \end{figure*} 
    
    Figure \ref{energycharg-fig} represents the energy charged ($e_{vnt}^{C}$) by electric vehicles before contingencies. It is noteworthy that when contingencies are not considered, the period in which electric vehicles are most charged is between $t=9$ and $t=16$, as it is a period with high availability of generation through renewable sources. When frequency reserve is considered, the system becomes more flexible, allowing electric vehicles to charge in other periods beyond the period in which renewable sources are available. Thus, these vehicles will provide frequency support in eventual failures of generating units.
    
    Related to the discharged energy by PEVs, only around the instant $t=22$ do the vehicle's discharge in all cases. Since it is a period during the night when the intermittent sources (Units $4$ and $5$) are no longer available for generation and still have relatively high demand, then PEVs can use stored energy to help meet demand.
    
\begin{figure*}[!h]
        \centering
        \includegraphics[scale=0.35]{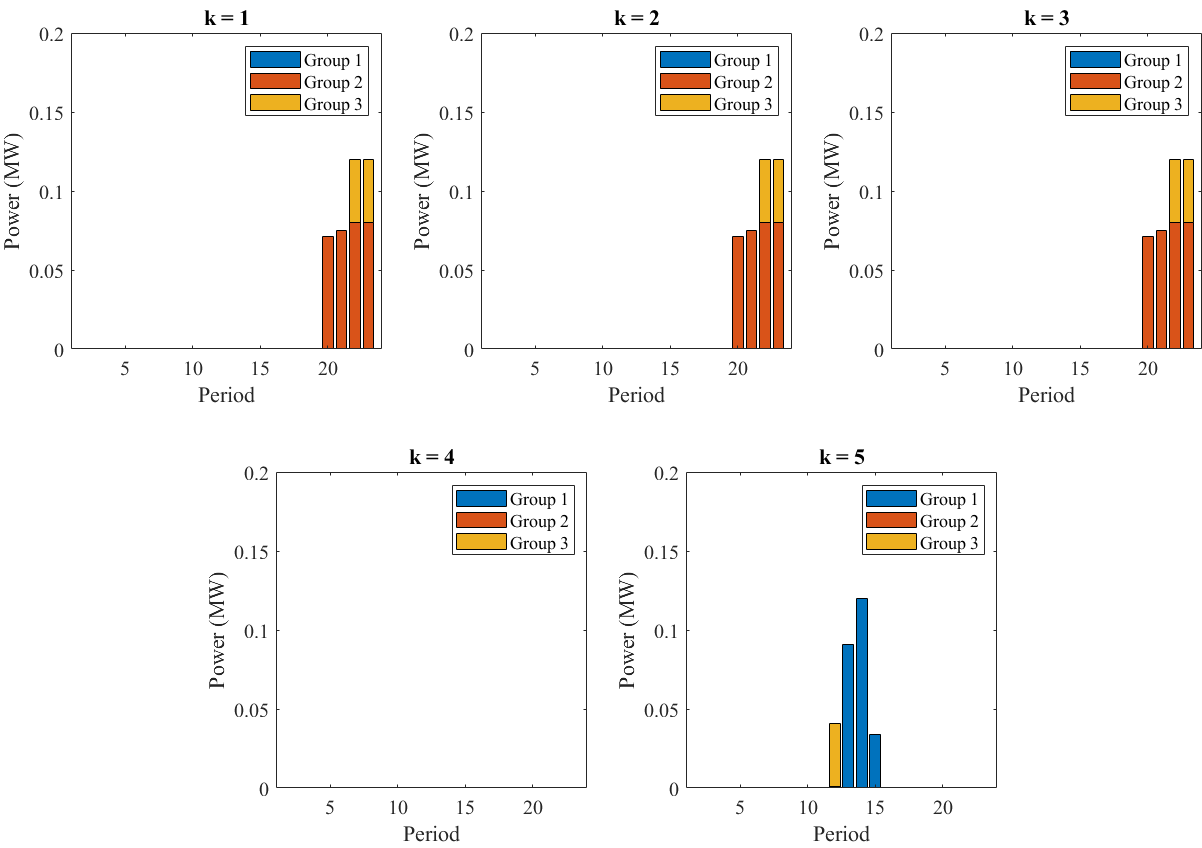}
        \caption{PFR provided by electric vehicles in case 3}
        \label{pfrbypevsc3-fig}
    \end{figure*}

    Figure \ref{pfrbypevsc3-fig} shows the PFR provided by electric vehicles ($p_{vntk}^{V,PR}$). The contingency $k=1$ represents the failure of the Unit $1$, the contingency $k=2$ the Unit $2$ failure and so on. 
    
    It is important to note that for the first three contingencies, the vehicles provide support in frequency in the periods referring to the night period when there is no availability of intermittent sources. At $k=4$, the PEVs did not provide PFR as the generating units could provide what was needed. In the $k=5$ contingency, the vehicles provided PFR during the daytime period, as this is the contingency that represents the unit's failure with the highest generation capacity (Unit 5). In addition, it is an intermittent unit that operates during the day.
    
    \begin{figure*}[!h]
        \centering
        \includegraphics[scale=0.35]{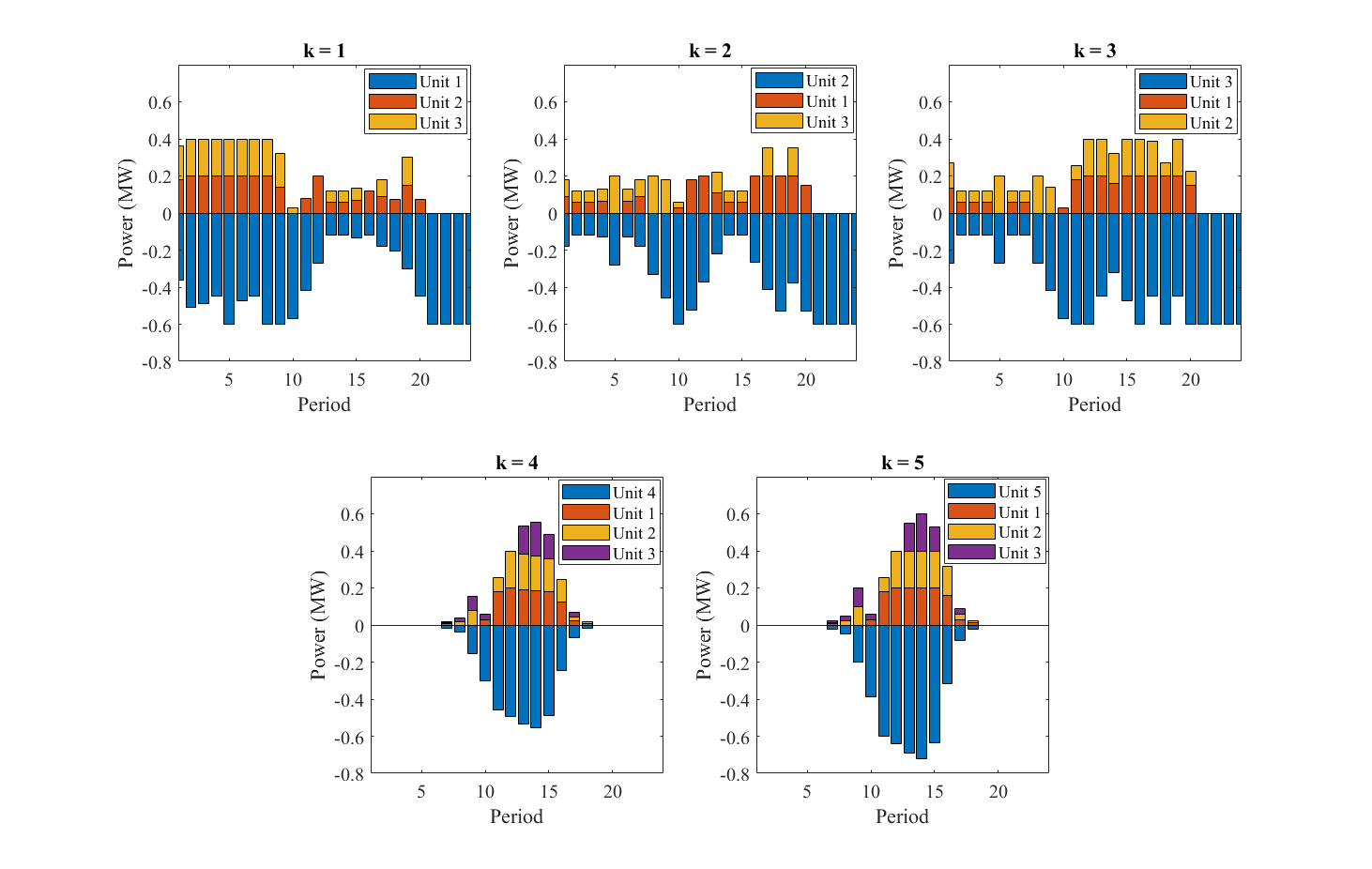}
        \caption{PFR provided by generating units in case 1}
        \label{prfbygenunitsc1-fig}
    \end{figure*}
    
    \begin{figure*}[!h]
        \centering
        \includegraphics[scale=0.35]{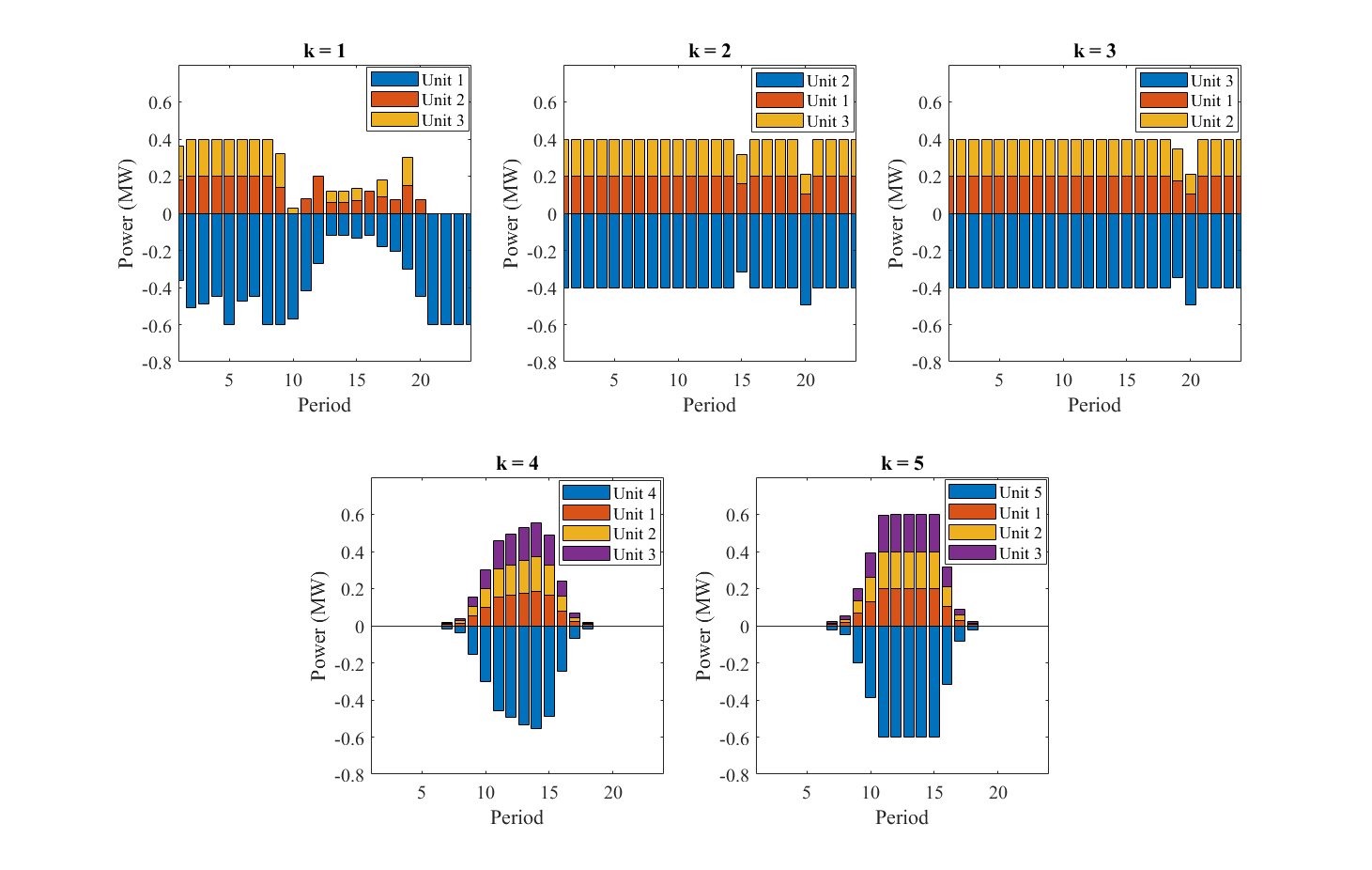}
        \caption{PFR provided by generating units in case 2}
        \label{prfbygenunitsc2-fig}
    \end{figure*}
    
    \begin{figure*}[!h]
        \centering
        \includegraphics[scale=0.35]{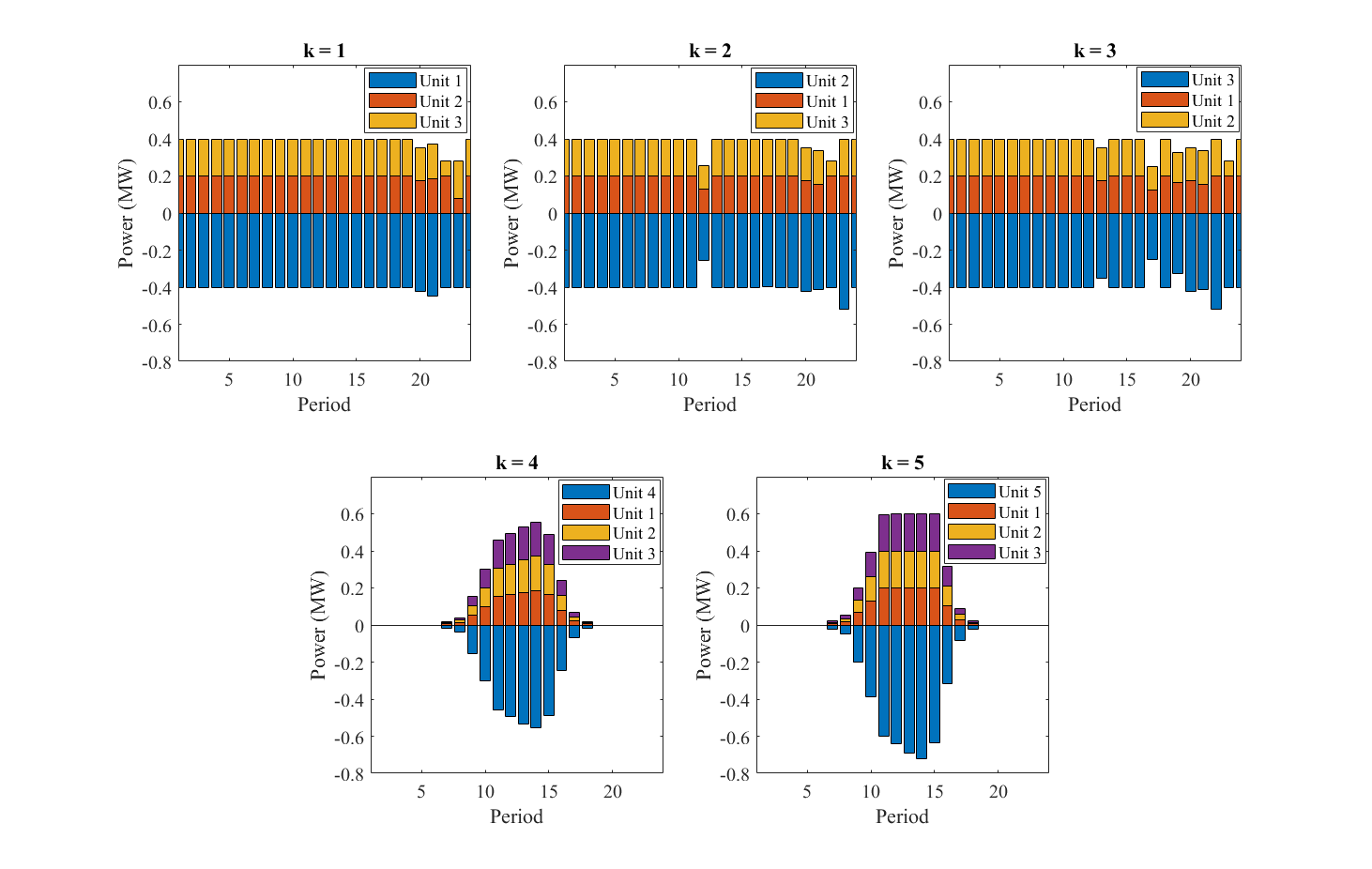}
        \caption{PFR provided by generating units in case 3}
        \label{prfbygenunitsc3-fig}
    \end{figure*}
    
    The Figures \ref{prfbygenunitsc1-fig}, \ref{prfbygenunitsc2-fig} and \ref{prfbygenunitsc3-fig}, show the PFR provided by each generating unit ($p_{gtk}^{PR}$) in each of the 3 cases studied and after each of the contingencies. The negative values in Figures \ref{prfbygenunitsc1-fig}, \ref{prfbygenunitsc2-fig} and \ref{prfbygenunitsc3-fig} represent the power that the missing generator was providing before the contingency, and the positive values represent the PFR provided by the other generating units. Thus, it is ideal that there is symmetry concerning the positive and negative values, which would mean that the system can supply the power of the missing generator. Thus, note that the case with the best performance was case 3, where frequency regulation is considered, and electric vehicles participate in this service.

\begin{figure*}[!h]
        \centering
        \includegraphics[scale=0.45]{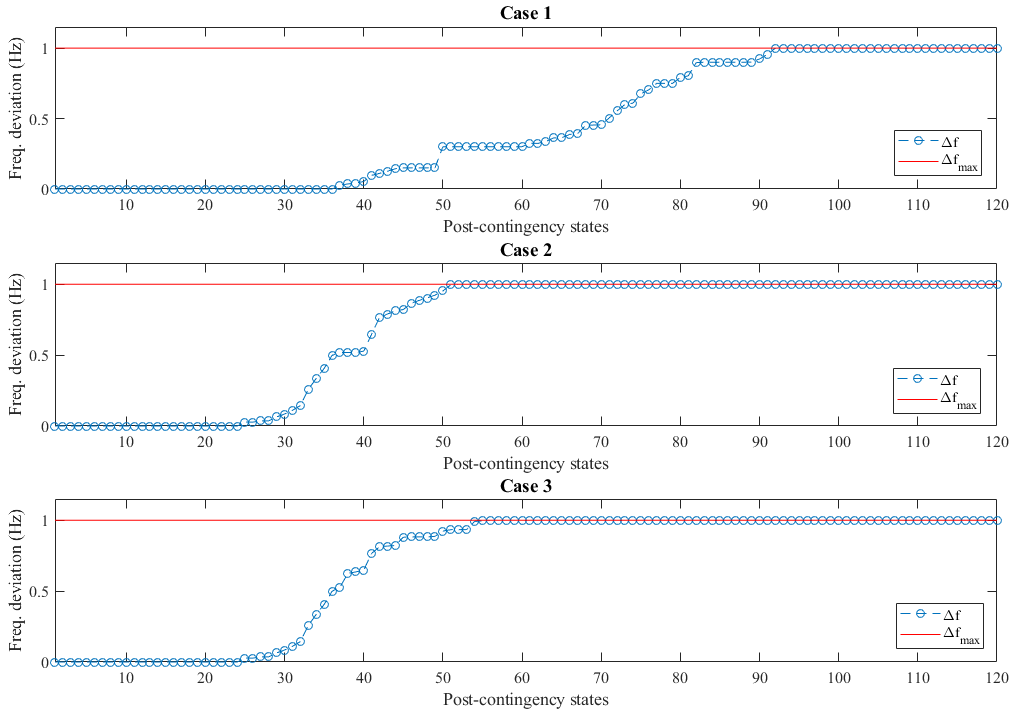}
        \caption{Frequency deviation in each period for each contingency considered for each analyzed case}    
        \label{energystinterm-fig}
    \end{figure*}

    
    The frequency deviation ($\Delta f_{tk}$) is shown in Figure \ref{energystinterm-fig}. The post-contingency states are shown on the horizontal axis, referring to each period for each contingency considered. This axis was reordered so that the frequency deviations are in ascending order, in order to obtain a better view of each case. Note that a greater frequency deviation was allowed to ensure that the unserved demand is as close to zero as possible. It is known that high-frequency variations can cause power quality problems, but as previously mentioned, a maximum frequency deviation of 1 Hz was considered. Note that Case 1 has a large number of post-contingency states with negligible frequency deviations. These post-contingency states correspond to contingencies of units with very small production.

\begin{figure*}[!h]
        \centering
        \includegraphics[scale=0.55]{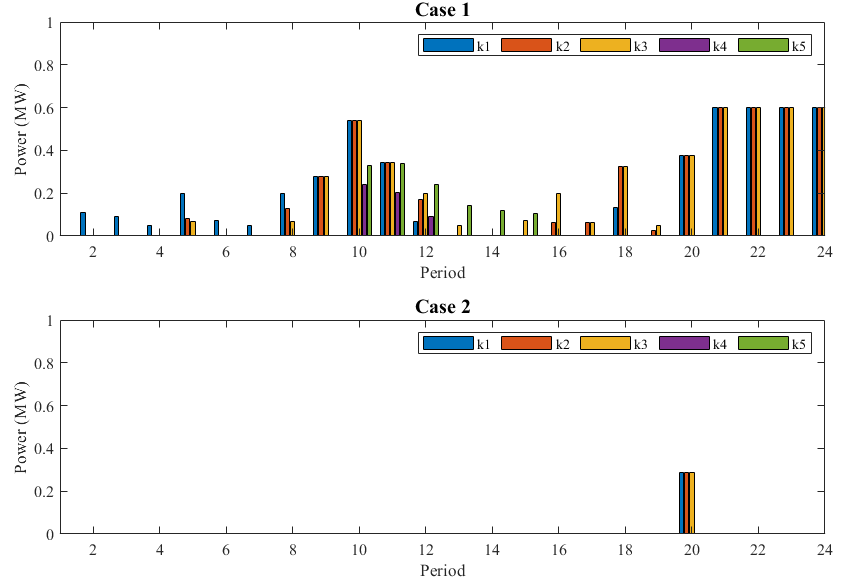}
        \caption{Unserved demand post-contingency}    
        \label{unserveddem-fig}
    \end{figure*}

    Finally, post-contingency unserved demand ($p_{dtk}^{UD,PR}$) is shown in Figure \ref{unserveddem-fig}. Note  there is not unserved demand in Case 3. In Case 2, there is unserved demand only in period $t=20$ after a contingency of one of the conventional generating units ($k=1$, $k=2$ and $k=3$). Note that, in this period, the intermittent units are no longer available for generation and there is a need to charge Group 2 of electric vehicles. Finally, observe that Case 1 has a significantly higher unserved demand than the rest of cases.
    
    A presentation of the day-ahead schedule costs for each case studied is shown in Tabele \ref{tab-costs}, it is possible to observe that the largest portion of the costs in case 1 is due to unserved demand after the contingencies, and in the other cases, this value decreases considerably.
    
    \begin{table*}[!h]
        \centering
        \caption{Costs for each case studied (R\$)}
        \hfill\par
        \begin{tabular}{ccccccc}
            \hline & $c_{t}^{P}$ & $c_{t}^{UD,PR}$ & $c_{t}^{\Delta f}$ & $cc_{t}^{V}$ & $cp_{t}^{V}$ & Total \\ \hline
            Case 1 &  14223.26&   165684.05& 5.41&       0.00&       0.00&     179.91 M\\
            Case 2 &  14591.18&     8669.48& 8.25&       0.00&       0.00&      23.27 M\\
            Case 3 &  14526.60&        0.00& 8.28&      33.55&     432.46&      15.00 M \\ \hline
        \end{tabular}
        \label{tab-costs}
    \end{table*}

\section{Conclusion}
    
    This work focused on analyzing the participation of PEVs in the day-ahead energy generating and reserve capacity scheduling, specifically, the participation of PEVs in the PFR of systems with high penetration of renewable energy sources. The proposed approach consists in a mathematical model that represents the day-ahead scheduling of a power system, which is formulated as a unit commitment problem.
    
    The model was applied to a case study based on electrical system of the Marco Zero Campus of the Federal University of Amapá. The results obtained make it possible to verify the importance of planning the system's operation considering the possibility of generating units failure concerning operating costs and system stability, thus making electrical power systems more flexible for the insertion of renewable sources. In addition, it was possible to estimate quantitatively the impact of the participation of PEVs in reducing the commitment of generating units that operate with a low capacity factor in the event of an unexpected generating units failure. It was possible to verify that there was a great improvement in the system operation when electric vehicles participated in different services in the system. This work also shows that in the scenario where there is a penalty for unserved demand at the Federal University of Amapá and where electric vehicles are available to provide ancillary services, the system's operating cost reduces considerably when these PEVs participate in frequency support.

    It is important to note that this work is considered a microgrid. In a large system, the contribution of electric vehicles in the system is expected to be more relevant.

\bibliographystyle{IEEEtran}
\bibliography{ref.bib}

\end{document}